\documentclass[12pt]{article}
\usepackage[utf8]{inputenc}
\usepackage{amsfonts,amsmath,amssymb,mathtools}
\usepackage[left=2cm,right=2cm,top=2cm,bottom=2cm,bindingoffset=0cm]{geometry}
\usepackage{graphicx}
\usepackage{cite}
\usepackage{authblk}
\usepackage{xcolor}
\usepackage{hyperref}

\newcommand\beq{\begin{equation}}
\newcommand\eeq{\end{equation}}
\newcommand{\pd}{\partial}
\newcommand{\mO}{\mathcal{O}}
\newcommand{\mN}{\mathcal{N}}
\newcommand{\const}{\text{const}}

\numberwithin{equation}{section}

\hypersetup{
    colorlinks=true,
    linkcolor=black,
    citecolor=blue,   
    urlcolor=blue,
}

\title{\bf Nonlinear dynamical Casimir effect at weak nonstationarity}
\author[1,2]{Dmitrii~A.~Trunin\thanks{\href{mailto:dmitriy.trunin@phystech.edu}{dmitriy.trunin@phystech.edu}}}
\affil[1]{Moscow Institute of Physics and Technology, 141700, Institutskiy pereulok, 9, Dolgoprudny, Russia}
\affil[2]{Institute for Theoretical and Experimental Physics, 117218, B. Cheremushkinskaya, 25, Moscow, Russia}
\date{\today}

\begin{document}

\maketitle

\begin{abstract}
We show that even small nonlinearities significantly affect particle production in the dynamical Casimir effect at large evolution times. To that end, we derive the effective Hamiltonian and resum leading loop corrections to the particle flux in a massless scalar field theory with time-dependent Dirichlet boundary conditions and quartic self-interaction. To perform the resummation, we assume small deviations from the equilibrium and employ a kind of rotating wave approximation. Besides that, we consider a quantum circuit analog of the dynamical Casimir effect, which is also essentially nonlinear. In both cases, loop contributions to the number of created particles are comparable to the tree-level values.
\end{abstract}

\setcounter{page}{0}
\thispagestyle{empty}
\newpage

\section{Introduction}
\label{sec:intro}

The dynamical Casimir effect (DCE) describes particle production by nonuniformly accelerated mirrors~\cite{Moore, DeWitt, Davies:1976, Davies:1977, Dodonov:1996, Lambrecht} and aptly illustrates many prominent features of the quantum field theory. On the one hand, this effect is similar to the celebrated Hawking~\cite{Hawking, Hawking:1974, Hawking:1976} and Unruh~\cite{Unruh, Fulling, Davies:1974} effects. Essentially, all these phenomena are related to a change in the system's quantum state due to the interactions with strong external fields; the only difference is the nature of the interactions and external fields. On the other hand, the DCE is amenable to a plethora of theoretical and experimental approaches, most of which are concisely described in reviews~\cite{Dodonov:2010, Dodonov:2020}. In particular, recently, this effect was experimentally measured in a system of superconducting quantum circuits~\cite{Wilson, Lahteenmaki, Nation}. The mentioned properties make the DCE a perfect model for the study of nonstationary quantum field theory. 

Most of the research on the DCE, including experimental implementations, focuses on a simplified two-dimensional model with a free massless scalar field and perfectly reflecting boundary conditions (we set $\hbar = c = 1$):
\beq \label{eq:DCE}
\left( \pd_t^2 - \pd_x^2 \right) \phi(t,x) = 0, \quad \phi[t, L(t)] = \phi[t, R(t)] = 0, \eeq
where mirror trajectories $L(t)$ and $R(t)$ are stationary in the asymptotic past and future, i.e., $L(t) \approx L_\pm + \beta_\pm t$ and $R(t) \approx L_\pm + \Lambda_\pm + \beta_\pm t$ as $t \to \pm \infty$.  Here, $\Lambda_\pm > 0$ is the distance between the mirrors, and $\beta_\pm$ is their velocity in the corresponding limits. Note that for physically meaningful trajectories, $L(t) < R(t)$ and $|\dot{L}(t)| < 1$, $|\dot{R}(t)| < 1$ for all $t$. Usually one also makes a Lorentz boost, shifts the origin and conveniently sets $L(t < 0) = 0$, $R(t < 0) = \Lambda_-$ without loss of generality.

Due to the nonstationarity of the model~\eqref{eq:DCE}, its Hamiltonian cannot be diagonalized once and forever, so the notion of a particle is different in the limits $t \to \pm \infty$. In other words, a natural mode decomposition of the free quantum field $\phi(t,x)$ is different in the asymptotic past and future:
\beq \phi(t,x) = \begin{cases} \sum_n \left[ a_n^\text{in} f_n^\text{in}(t,x) + H.c. \right], \quad &\text{as} \quad t \to -\infty, \\ \sum_n \left[ a_n^\text{out} f_n^\text{out}(t,x) + H.c. \right], \quad &\text{as} \quad t \to +\infty, \end{cases} \eeq
where functions $f_n^\text{in}$, $f_n^\text{out}$ solve the free equations of motion and form a complete basis with respect to a proper inner product. In general, the modes and creation/annihilation operators that diagonalize the free Hamiltonian in the asymptotic past (in-region) and future (out-region) are related by a generalized Bogoliubov (or canonical) transformation~\cite{Birrell, Fulling:1989, Grib}:
\beq \begin{aligned}\
f_n^\text{out} &= \sum_k \left[ \alpha_{nk}^* f_k^\text{in} - \beta_{nk} (f_k^\text{in})^* \right], \\
a_n^\text{out} &= \sum_k \left[ \alpha_{kn} a_k^\text{in} + \beta_{kn}^* (a_k^\text{in})^\dag \right],
\end{aligned} \eeq
with nonzero Bogoliubov coefficients $\alpha_{nk}$ and $\beta_{nk}$. Hence, the number of the out-particles ($f_n^\text{out}$ modes) created during the nonstationary evolution, in the Heisenberg picture, is given by the following expression (see Appendix~\ref{sec:N-phys} for an alternative derivation):
\beq \label{eq:N}
\mN_n = \langle in | (a_n^\text{out})^\dag a_n^\text{out} | in \rangle = \sum_k | \beta_{kn}|^2 + \sum_{k,l} \left( \alpha_{kn} \alpha_{ln}^* + \beta_{kn} \beta_{ln}^* \right) n_{kl}  + \sum_{k,l} \alpha_{kn} \beta_{ln} \kappa_{kl} + \sum_{k,l} \alpha_{kn}^* \beta_{ln}^* \kappa_{kl}^*. \eeq
Here, we introduce the short notations for the level density $n_{kl} = \langle in | (a_k^\text{in})^\dag a_l^\text{in} | in \rangle$ and anomalous quantum average (correlated pair density) $\kappa_{kl} = \langle in | a_k^\text{in} a_l^\text{in} | in \rangle$ and schematically denote the initial quantum state (possibly mixed) as $| in \rangle$.

In the Gaussian approximation, i.e., in the absence of interactions and nonlinearities, particle creation in the DCE is related only to mixing of positive- and negative-frequency modes. Indeed, assume that the initial quantum state of the system coincides with the vacuum, $a_n^\text{in} | in \rangle = 0$ for all~$n$. In this case, both $n_{kl} = 0$ and $\kappa_{kl} = 0$, so the identity~\eqref{eq:N} significantly simplifies:
\beq \label{eq:N-tree}
\mN_n^\text{free} = \sum_k |\beta_{kn}|^2. \eeq
This identity also approximately holds at low temperatures: in this case, $n_{kl}$ is exponentially suppressed and $\kappa_{kl}$ is zero, so they give negligible contributions to the right-hand side of~\eqref{eq:N}. Thus, on the tree level, particle creation is fully determined by the Bogoliubov coefficient $\beta_{kn}$. For this reason, identity~\eqref{eq:N-tree} is widely considered as the essence of the DCE, and the vast majority of literature --- including seminal papers~\cite{Moore, DeWitt, Davies:1976, Davies:1977, Dodonov:1996, Lambrecht}, reviews~\cite{Dodonov:2010, Dodonov:2020, Nation}, and textbooks~\cite{Birrell, Fulling:1989, Grib} --- is purely dedicated to the calculation of $\mN_n^\text{free}$ and derivative quantities.

Nevertheless, real-world systems are rarely free from interactions and nonlinearities. In such systems, approximation~\eqref{eq:N-tree} is not applicable even if the initial values of $n_{kl}$ and $\kappa_{kl}$ are small. In fact, in nonstationary situations, these quantities can receive nonnegligible loop corrections\footnote{Of course, in nonstationary systems, loop corrections should be calculated using a nonstationary (Schwinger-Keldysh) diagrammatic technique~\cite{Schwinger, Keldysh, Kamenev, Rammer, Landau:vol10, Berges, Arseev}.}:
\begin{align}
\label{eq:n}
n_{kl}(t,t_0) &= \langle in | U^\dag(t, t_0) (a_k^\text{in})^\dag a_l^\text{in} U(t, t_0) | in \rangle, \\
\label{eq:k}
\kappa_{kl}(t,t_0) &= \langle in | U^\dag(t, t_0) a_k^\text{in} a_l^\text{in} U(t, t_0) | in \rangle,
\end{align}
where $U(t,t_0)$ denotes the evolution operator in the interaction picture, and $t_0$, $t$ set the moments when the interaction term is adiabatically switched on and off. In this case, identity~\eqref{eq:N} does not reduce to~\eqref{eq:N-tree}.

Furthermore, loop corrections can be significant even if interactions are apparently weak. As was recently discovered, in many nonstationary interacting systems, $n_{kl}$ and/or $\kappa_{kl}$ receive secularly growing loop corrections, i.e., $n_{kl}^{(n)} \sim (\lambda t)^{a_n}$ and $\kappa_{kl}^{(n)} \sim (\lambda t)^{b_n}$ with some constant integer powers $a_n > 0$ and $b_n > 0$ in every order of the perturbation theory in $\lambda$. Such corrections indefinitely grow in the limit $t \to \infty$ and are not suppressed even if the coupling constant goes to zero, $\lambda \to 0$. For instance, the secular growth of loop corrections was observed in the de~Sitter~\cite{Akhmedov:dS, Krotov, Polyakov, Bascone, Pavlenko, Moschella} and Rindler~\cite{Bazarov} spaces, strong electric~\cite{Musaev, Akhmedov:Et, Akhmedov:Ex} and scalar~\cite{Diatlyk-1, Lanina, Diatlyk-2} fields, gravitational collapse~\cite{Akhmedov:H}, and nonstationary quantum mechanics~\cite{Trunin-1, Trunin-QM}. A short review of the origin and consequences of the secularly growing loops can be found in~\cite{Akhmedov:2021}.

The DCE was also recently shown to possess secularly growing loop corrections~\cite{Alexeev, Akopyan}. Unfortunately, this analysis was restricted to the first few loops. At the same time, due to the secular growth, high and low order loop corrections in this model are comparable at large evolution times. Therefore, a definitive conclusion about the destiny of $n_{kl}$ and $\kappa_{kl}$ can be made only after a resummation of the leading contributions from all loops. In other words, the number of created particles~\eqref{eq:N} and other observable quantities in the nonlinear DCE have physical meaning only nonperturbatively.

In this paper, we estimate a nonperturbative contribution to the number of created particles in the nonlinear scalar DCE. For illustrative purposes, in most of the paper, we consider a simple nonlinear generalization of the model~\eqref{eq:DCE} with a quartic potential:
\beq \label{eq:DCE-nlin}
\left( \pd_t^2 - \pd_x^2 \right) \phi(t,x) = \lambda \phi^3(t,x), \quad \phi[t, L(t)] = \phi[t, R(t)] = 0, \eeq
restrict ourselves to small deviations from stationarity and assume that the initial state of the field coincides with the vacuum. This approximation simplifies the calculation of the Bogoliubov coefficients and allows us to resum the perturbative series of loop corrections to $n_{kl}$ and $\kappa_{kl}$. As a result, we obtain that the nonperturbative loop contributions to~\eqref{eq:N} in the approximation mentioned above are comparable to the tree-level approximation~\eqref{eq:N-tree} at large evolution times --- even when the coupling constant $\lambda \to 0$.

In addition, we extend these calculations to a more physically relevant model:
\beq \label{eq:DCE-squid}
\left[ \frac{1}{v^2(t,x)} \pd_t^2 - \pd_x^2 \right] \phi(t,x) = -\lambda \pd_x \left[ \pd_x \phi(t,x) \right]^3, \quad \phi(t,0) = \phi(t, \Lambda) = 0, \eeq
which naturally emerges in a Josephson metamaterial implementation of the DCE~\cite{Lahteenmaki} due to the small nonlinearity of superconducting quantum interference devices (see Appendix~\ref{sec:Josephson} for the details). For small deviations from stationarity, i.e., for relatively small variations of the light speed $v(t,x)$, this model is qualitatively equivalent to a simplified model~\eqref{eq:DCE-nlin}. In this case, the resummed loop contribution to the created particle number is also significant at large evolution times. 

This paper is organized as follows. In Sec.~\ref{sec:Bogoliubov}, we discuss the quantization of a free scalar field and calculate the Bogoliubov coefficients in the model~\eqref{eq:DCE}. In Sec.~\ref{sec:H}, we derive the effective Hamiltonian of the nonlinear model~\eqref{eq:DCE-nlin}. In Sec.~\ref{sec:QM}, we calculate level density and anomalous quantum average in the assumption that almost all created particles populate the fundamental (lowest-energy) mode. In this case, the effective Hamiltonian is essentially quantum mechanical. In Sec.~\ref{sec:large-N}, we consider a large~$N$ model, which is qualitatively equivalent to the model~\eqref{eq:DCE-nlin}, and confirm that at small deviations from stationarity, the fundamental mode is the most populous one. Thus we justify the assumption of Sec.~\ref{sec:QM}. In Sec.~\ref{sec:circuit}, we repeat these calculations for a physically motivated model~\eqref{eq:DCE-squid}. Finally, we discuss the results and conclude in Sec.~\ref{sec:discussion}. Besides, we explain the physical meaning of $\mN_n$ in Appendix~\ref{sec:N-phys}, present the full expression for the interacting Hamiltonian of the model~\eqref{eq:DCE-nlin} in Appendix~\ref{sec:H-full} and derive the model~\eqref{eq:DCE-squid} from the  Hamiltonian of a Josephson metamaterial in Appendix~\ref{sec:Josephson}.

\section{Field quantization and Bogoliubov coefficients}
\label{sec:Bogoliubov}

Consider a quantization of the free model~\eqref{eq:DCE} with $L(t < 0) = 0$ and $R(t < 0) = \Lambda_-$, i.e., initially resting mirrors. As usual, we expand the operator of the quantized field in the mode functions:
\beq \label{eq:phi}
\phi(t,x) = \sum_{n=1}^\infty \left[ a_n^\text{in} f_n^\text{in}(t,x) + H.c. \right]. \eeq
Here, operators $a_n^\text{in}$ and $(a_n^\text{in})^\dag$ satisfy the standard bosonic commutation relations, and mode functions $f_n^\text{in}$ are expressed in terms of auxiliary functions $G(z)$ and $F(z)$:
\beq \label{eq:in}
f_n^\text{in}(t,x) = \frac{i}{\sqrt{4 \pi n}} \left[ e^{-i \pi n G(t + x)} - e^{- i \pi n F(t - x)} \right], \eeq
which solve the generalized Moore's equations~\cite{Moore}:
\beq \label{eq:Moore}
G\left[ t + L(t) \right] - F\left[ t - L(t) \right] = 0, \qquad G\left[ t + R(t) \right] - F\left[ t - R(t) \right] = 2 \eeq
and comply the initial conditions $G(z \le \Lambda_-) = F(z \le 0) = z/\Lambda_-$. Note that due to stationarity of the motion of mirrors in the asymptotic future, functions $G(z)$ and $F(z)$ periodically grow as $z \to +\infty$: $G(z + \Delta z_G) = G(z) + 2$ and $F(z + \Delta z_F) = F(z) + 2$ with $\Delta z_G = \frac{2 \Lambda_+}{1 - \beta_+}$ and $\Delta z_F = \frac{2 \Lambda_+}{1 + \beta_+}$. This property can be easily inferred from the geometric method for constructing modes~\cite{Akopyan, Cole, Meplan, Li}. Moreover, in this limit, the first Moore's equation implies a simple relation:
\beq F(z) = G\big( \tilde{z}(z) \big), \quad \tilde{z}(z) = \frac{1 + \beta_+}{1 - \beta_+} z + \frac{2 L_+}{1 - \beta_+}, \quad \text{as} \quad z \to +\infty. \eeq

Let us return to mode functions~\eqref{eq:in}. First, they can be straightforwardly shown to solve the free equation of motion and form a complete orthonormal basis with respect to the Klein-Gordon inner product:
\beq \label{eq:inner}
(u, v) = -i \int_{L(t)}^{R(t)} \left[ u \pd_t v^* - v^* \pd_t u \right] dx. \eeq
Second, these functions have definite positive frequency with respect to the Killing vector $\xi = \pd_t$:
\beq f_n^\text{in}(t,x) = \frac{1}{\sqrt{\pi n}} \exp\left( -i \frac{\pi n t}{\Lambda_-} \right) \sin\left( \frac{\pi n x}{\Lambda_-} \right), \quad \text{as} \quad t < 0, \eeq
and diagonalize the free Hamiltonian at the past infinity:
\beq \label{eq:H-free-def}
H_\text{free}(t) = \int_{L(t)}^{R(t)} dx \left[ \frac{1}{2} (\pd_t \phi)^2 + \frac{1}{2} (\pd_x \phi)^2 \right] = \sum_{n=1}^\infty \frac{\pi n}{\Lambda_-} \left[ (a_n^\text{in})^\dag a_n^\text{in} + \frac{1}{2} \right], \quad \text{as} \quad t < 0. \eeq
Due to these reasons, functions~\eqref{eq:in} are usually referred to as in-modes.

At the same time, at the future infinity in-modes contain both positive and negative frequencies:
\beq \label{eq:in-future}
f_n^\text{in}(t,x) \to \sum_{k=1}^\infty \frac{1}{\sqrt{\pi k}} \left[ \alpha_{nk} \exp\left( -i \frac{\pi k \tilde{t}}{\tilde{\Lambda}} \right) + \beta_{nk} \exp\left( i \frac{\pi k \tilde{t}}{\tilde{\Lambda}} \right) \right] \sin\left( \frac{\pi k \tilde{x}}{\tilde{\Lambda}} \right), \quad \text{as} \quad t \to +\infty, \eeq
where we change to the coordinates $\tilde{t} = \gamma_+ (t - \beta_+ x)$, $\tilde{x} =\gamma_+ (x - \beta_+ t - L_+)$ and denote $\tilde{\Lambda} = \gamma_+ \Lambda_+$, $\gamma_+ = 1/\sqrt{1 - \beta_+^2}$ for shortness. The coefficients in the decomposition~\eqref{eq:in-future} are nothing but the Bogoliubov coefficients determined by the inner products~\eqref{eq:inner}:
\beq \label{eq:B-def}
\alpha_{nk} = \left( f_n^\text{in}, f_k^\text{out} \right), \quad \beta_{nk} = -\left( f_n^\text{in}, (f_k^\text{out})^* \right), \eeq
of in-modes~\eqref{eq:in} and out-modes:
\beq f_n^\text{out}(t,x) \to \frac{1}{\sqrt{\pi n}} \exp\left( -i \frac{\pi n \tilde{t}}{\tilde{\Lambda}} \right) \sin\left( \frac{\pi n \tilde{x}}{\tilde{\Lambda}} \right), \quad \text{as} \quad t \to +\infty. \eeq
For arbitrary mirror motion, functions $G(z)$, $F(z)$ and the Bogoliubov coefficients are very difficult to compute. Nevertheless, this computation significantly simplifies if we restrict ourselves to weak deviations from stationarity, i.e., consider trajectories of the form $L(t) = \epsilon l(t)$, $R(t) = \Lambda_- + \epsilon r(t)$, $\epsilon \ll 1$, and exclude motions that induce constructive interference. In this case, $G(z)$ and inverse function $g(y) = G^{-1}(y)$ are approximately linear: $G(z) \approx \frac{2}{\Delta z_G} z + \mO(\epsilon)$, $g(y) \approx \frac{\Delta z_G}{2} y + \mO(\epsilon)$. Keeping in mind that these functions periodically grow with $z$ or $y$, we expand their derivatives into the Fourier series:
\beq \label{eq:Fourier} \begin{aligned}
G'(z) &= \sum_{n = -\infty}^\infty G_n e^{i \frac{2 \pi}{\Delta z_G} n z}, \quad & G_n &= \frac{1}{\Delta z_G} \int_0^{\Delta z_G} G'(z) e^{-i \frac{2 \pi}{\Delta z_G} n z} dz, \\ g'(y) &= \sum_{n = -\infty}^\infty g_n e^{i \pi n y}, \quad & g_n &= \frac{1}{2} \int_0^2 g'(y) e^{- i \pi n y} dy,
\end{aligned} \eeq
and find high order Fourier coefficients to be small, $G_n \sim \epsilon$ and $g_n \sim \epsilon$ for all $n \neq 0$. In addition, these coefficients are symmetric, $G_{-n} = G_n^*$ and $g_{-n} = g_n^*$, since the original functions are real. These properties allow us to expand complex exponents from~\eqref{eq:in} into simple plane waves:
\beq \frac{1}{\Delta z_G} \int_0^{\Delta z_G} e^{-i \pi n G(z)} e^{-i \frac{2 \pi}{\Delta z_G} k z} dz 
= \delta_{n,-k} + \frac{g_{n+k}}{g_0} \frac{n}{n+k} \left( 1 - \delta_{n,-k} \right) + \mO\left( \epsilon^2 \right), \eeq
hence,
\beq e^{-i \pi n G(z)} \approx e^{-i \frac{2 \pi}{\Delta z_G} n z} + \sum_{k \neq n} \frac{g_{n-k}}{g_0} \frac{n}{n-k} e^{-i \frac{2 \pi}{\Delta z_G} k z}. \eeq
Combining this identity with the relation~\eqref{eq:Moore}, changing to the $(\tilde{t}, \tilde{x})$ coordinates and calculating inner products~\eqref{eq:B-def}, we straightforwardly obtain the Bogoliubov coefficients (we single out the Kronecker delta $\delta_{n,k}$ to make the $\epsilon$-expansion explicit):
\beq \label{eq:B}
\begin{aligned} \alpha_{nk} &= \delta_{n,k} + \frac{g_{n-k}}{g_0} \frac{\sqrt{n k}}{n-k} \left( 1 - \delta_{n,k} \right) + \mO\left(\epsilon^2\right), \\
\beta_{nk} &= -\frac{g_{n+k}}{g_0} \frac{\sqrt{n k}}{n+k} + \mO\left(\epsilon^2\right).
\end{aligned} \eeq
We emphasize that these approximate identities are valid only for relatively small frequencies, $n \ll 1/\epsilon$, where particle wavelength is much larger than the characteristic mirror displacement, $\bar{\lambda} \sim \tilde{\Lambda} / \pi n \gg \epsilon \tilde{\Lambda}$. We also expect that at high frequencies, interactions between the field and the mirror can be approximated by a quasistationary process (at least if we assume no constructive interference). Hence, the corrections to the stationary Bogoliubov coefficients (i.e., identity transformation) in this case are approximately zero, $\alpha_{n \neq k} \approx 0$ and $\beta_{nk} \approx 0$ for $n \gg 1/\epsilon$ or $k \gg 1/\epsilon$.

We remind that identities~\eqref{eq:B} are valid only for weakly nonstationary motions. Notable examples of such a motion include ``broken'' trajectories with a small final velocity $\epsilon$:
\beq \label{eq:broken}
\begin{gathered}
L(t) = \epsilon t \theta(t), \quad R(t) = \Lambda + \epsilon t \theta(t), \\
g_0 = \frac{\Lambda}{1 - \epsilon}, \quad g_{n \neq 0} = - \frac{2 \epsilon \Lambda}{1 - \epsilon} \frac{1 - (-1)^n}{2 i \pi n},
\end{gathered} \eeq
or resonant oscillations with destructive interference\footnote{We emphasize that this is a very special case of resonant oscillations; in fact, the interference is constructive for the majority of oscillation amplitudes, frequencies, and dephasing angles~\cite{Dalvit:1997, Dalvit:1998}.}, e.g.:
\beq \label{eq:resonant}
\begin{gathered}
L(t) = \epsilon \Lambda \sin\left( \frac{\pi q t}{\Lambda} \right) \theta(t), \quad R(t) = \Lambda + \epsilon \Lambda \sin\left( \frac{\pi q t}{\Lambda} \right) \theta(t), \\
g_0 = \Lambda, \quad g_{\pm q} \approx \frac{q}{2} \pi \epsilon \Lambda, \quad g_{n \neq 0, \pm q} \approx \epsilon \Lambda \frac{2 i q n}{n^2 - q^2} \frac{1 - (-1)^n}{2},
\end{gathered} \eeq
where $q = 2, 4, 6, \cdots$. The derivation\footnote{In comparison, the direct calculation of $\beta_{nk}$ for ``broken'' trajectories~\eqref{eq:broken} yields the following expression:
$$ \beta_{nk} = \frac{2 \epsilon}{i \pi} \frac{n}{(n+k)^2 - (\epsilon k)^2} \frac{1 - (-1)^{n+k} e^{i \pi \epsilon k}}{2} \sqrt{\frac{k}{n}} = -\frac{g_{n+k}}{g_0} \frac{\sqrt{nk}}{n+k} + \mO\left(\epsilon^2\right), $$
which confirms the validity of approximations~\eqref{eq:B}. The apparently troubling phase factor $e^{i \pi \epsilon k}$ is significant only at very large mode numbers, $k \sim 1/\epsilon$, where its prefactor is itself proportional to $\mO\left(\epsilon^2\right)$.} of coefficients $g_n$ and $\beta_{nk}$ for the motions~\eqref{eq:broken} and~\eqref{eq:resonant} can be found in~\cite{Akopyan, Dalvit:1998}.

In strongly nonstationary situations, the relations between the mirror trajectories, functions $G(z)$ and $g(y)$, the Bogoliubov coefficients, and the number of created particles are more complicated. The most notable examples of such situations include resonant oscillations with a constructive interference~\cite{Dodonov:1996, Lambrecht, Dalvit:1997, Dalvit:1998, Schutzhold, Wu, Dodonov:1998} and an asymptotic light-like mirror motion~\cite{Davies:1976, Davies:1977, Castagnino, Walker, Carlitz, Obadia-1, Obadia-2, Obadia-3, Good}. Nevertheless, recall that in the present paper, we confine our study to weakly nonstationary motions, for which approximate identities~\eqref{eq:B} can be safely used.

\section{Effective Hamiltonian}
\label{sec:H}

The full Hamiltonian of the nonlinear DCE, model~\eqref{eq:DCE-nlin}, is very complex. Therefore, we need a reasonable approximation to simplify the calculations and estimate the number of created particles. First, we assume small deviations from stationarity, i.e., $\beta_{nk} \sim \alpha_{n \neq k} \sim \mO(\epsilon)$ with $\epsilon \ll 1$. In the leading order in $\epsilon$, this approximation forbids almost all processes that violate the energy conservation law. Second, we consider small coupling constants and large evolution times\footnote{This limit is equivalent to the limit $\lambda \to 0$, $\tilde{t} \to \infty$, $\lambda \tilde{\Lambda} \tilde{t} = \const$, if the velocity of the mirrors does not approach the speed of light in the future infinity.}, $\lambda \to 0$, $t \to \infty$, $\lambda \Lambda_+ t = \const$, i.e., keep only the leading secularly growing contributions to the evolution operator. This limit is similar to the rotating-wave approximation in quantum optics~\cite{Walls, Irish, Law, Recamier} and suppresses rapidly oscillating parts of the effective Hamiltonian. We also note that in this limit, the Bogoliubov coefficients are approximately constant. Both these approximations are inspired by a quantum mechanical toy model of nonstationary particle production~\cite{Trunin-QM, Dodonov:1993, Dodonov:1995}.

First of all, let us calculate the free Hamiltonian in the suggested limit. Substituting mode decomposition~\eqref{eq:in-future} into identity~\eqref{eq:H-free-def}, rotating to the $(\tilde{t}, \tilde{x})$ coordinates, integrating over $\tilde{x}$ and neglecting rapidly oscillating and constant contributions, we obtain the following approximate identity:
\beq \label{eq:H-free}
\begin{aligned}
H_\text{free} &= \sum_{m,n,k=1}^\infty \frac{\omega_k}{2} \Big[ \left(\alpha_{mk}^* \alpha_{nk} + \beta_{mk}^* \beta_{nk} \right) a_m^\dag a_n + \left( \alpha_{mk} \beta_{nk} + \beta_{mk} \alpha_{nk} \right) a_m a_n + H.c. \Big] \\
&\approx \sum_{n=1}^\infty \frac{1}{2} \omega_n  a_n^\dag a_n - \sum_{n,k=1}^\infty \frac{\omega_n + \omega_k}{2} \frac{g_{n+k}}{g_0} \frac{\sqrt{nk}}{n+k} a_n a_k + H.c. + \mO\left( \epsilon^2 \right),
\end{aligned} \eeq
as $t \to +\infty$. Here, we denote the frequency of the $f_n^\text{out}$ mode as $\omega_n = \pi n / \tilde{\Lambda}$. In the last line, we also substitute the approximate Bogoliubov coefficients~\eqref{eq:B} and keep only two leading terms in the $\epsilon$-expansion.

Note that the Hamiltonian of the massive variant of the model~\eqref{eq:DCE} also has the form~\eqref{eq:H-free} with a frequency $\omega_n = \sqrt{\left( \frac{\pi n}{\tilde{\Lambda}} \right)^2 + m^2} \approx \frac{\pi n}{\tilde{\Lambda}} + \frac{\tilde{\Lambda}}{2 \pi n} m^2$, as $m \to 0$. At the same time, the Moore's approach~\cite{Moore}, which we used to calculate the Bogoliubov coefficients in Sec.~\ref{sec:Bogoliubov}, does not work in the massive model due to the lack of conformal invariance. Hence, in the massive theory, approximation~\eqref{eq:B} and the second identity in~\eqref{eq:H-free} are valid only at relatively small evolution times, $\tilde{t} \ll 1/m^2 \tilde{\Lambda}$.

Now we employ the same method to calculate the interacting Hamiltonian at the future infinity:
\beq \label{eq:H}
H_\text{int} = \delta H_\text{free} + H_\text{int}^{(0)} + H_\text{int}^{(1)} + \mO\left( \epsilon^2 \right), \quad \text{as} \quad t \to +\infty,
\eeq
where $H_\text{int}^{(0)}$ and $H_\text{int}^{(1)}$ are the normal-ordered quartic interaction terms proportional to $\epsilon^0$ and $\epsilon^1$, respectively:
\beq
\label{eq:H-0}
H_\text{int}^{(0)} = \frac{\lambda \tilde{\Lambda}}{32 \pi^2} \sum_{k,l,m,n=1}^\infty \left[ \frac{3 \delta_{k+l,m+n} + 6 \delta_{k,m} \delta_{l,n}}{\sqrt{klmn}} a_k^\dag a_l^\dag a_m a_n - \frac{4 \delta_{k,l+m+n}}{\sqrt{klmn}} a_k^\dag a_l a_m a_n + H.c. \right],
\eeq
\beq \label{eq:H-1}
\begin{aligned}
H_\text{int}^{(1)} = \frac{\lambda \tilde{\Lambda}}{32 \pi^2 g_0} \sum_{k,l,m,n=1}^\infty \frac{1}{\sqrt{klmn}} \Big[ &\left( 3 g_{-k-l+m+n} + 12 \delta_{k,m} g_{-l+n} \right) \left( 1 - \delta_{k+l,m+n} \right) a_k^\dag a_l^\dag a_m a_n \\ &- \left( 4 g_{-k+l+m+n} + 12 \delta_{k,l} g_{m+n} \right) \left( 1 - \delta_{k,l+m+n} \right) a_k^\dag a_l a_m a_n \\ &+ g_{k+l+m+n} \, a_k a_l a_m a_n + H.c. \Big],
\end{aligned}
\eeq
and $\delta H_\text{free}$ is the normal-ordered quadratic term, which can be absorbed into the renormalization of frequency in the free Hamiltonian~\eqref{eq:H-free}:
\beq \delta H_\text{free} = \sum_{n=1}^\infty \frac{\tilde{\Lambda}}{2 \pi} \delta m^2 \left[ \frac{1}{n} a_n^\dag a_n - \sum_{k=1}^\infty \frac{g_{n+k}}{g_0} \frac{1}{\sqrt{nk}} a_n a_k - \sum_{k=1}^\infty \frac{g_{n+k}^*}{g_0} \frac{1}{\sqrt{nk}} a_n^\dag a_k^\dag \right] + \mO\left( \epsilon^2 \right). \eeq
In the last identity, we single out the correction to the physical mass (which is, essentially, the one-loop correction):
\beq \label{eq:dm}
\delta m^2 \approx \frac{3 \lambda}{2 \pi} \sum_{n=1}^{n_\text{UV}} \frac{1}{n} = \frac{3 \lambda}{2 \pi} \log \left( n_\text{UV} \right), \eeq
and introduce the ultraviolet cutoff $n_\text{UV}$. In what follows, we assume that this correction is canceled by the corresponding tree-level counterterm\footnote{However, note that the loop correction is positive for $\lambda > 0$, i.e., the physical mass can be zero only if the bare mass is tachyonic, $m_0^2 < 0$.}. Furthermore, we believe that this assumption does not affect our calculations. First, they are devoted to infrared rather than ultraviolet contributions. Second, we are interested in the evolution of the quantum state, which is not directly related to the mass of the field. For these reasons, in the following sections, we assume zero physical mass.

Note that the Kronecker deltas in the $H_\text{int}^{(0)}$ establish energy conservation in the scattering ($a^\dag a^\dag a a$) and decay ($a^\dag a a a$ and $a^\dag a^\dag a^\dag a$) processes. This is a consequence of the emergent time translation symmetry in the limit $\epsilon \to 0$. Due to the same reason, there are no substantial loop corrections in this limit. Conversely, higher-order parts of the interacting Hamiltonian violate the energy conservation law\footnote{Essentially, this violation is related to the energy exchange with the external world, which becomes possible due to the presence of mirrors.} and allow usually forbidden processes. For instance, the $a a a a$ term in the $H_\text{int}^{(1)}$ describes a simultaneous production of four correlated particles. However, in what follows, we restrict ourselves to the leading nontrivial order in $\epsilon$, i.e., assume that the ``forbidden'' processes occur only once or twice during the evolution of the system.

We also remind that in this section, we truncate the $\epsilon$-expansion of the interacting Hamiltonian~\eqref{eq:H} after the first order term. The full expression, which is expressed through the exact Bogoliubov coefficients and do not require the limit $\epsilon \to 0$, can be found in Appendix~\ref{sec:H-full}.

\section{Reduction to quantum mechanics}
\label{sec:QM}

As explained in the previous section, the limit $\epsilon \to 0$ establishes an approximate energy conservation, i.e., restricts the energy exchange with the external world. Hence, we expect this limit to imply a small number of created particles. Furthermore, we can expect that created particles mainly populate the most easily excited, lowest-energy mode with $n=1$. In this case, we can ignore all modes with $n > 1$ and reduce the interacting Hamiltonian~\eqref{eq:H} to a quantum-mechanical one:
\beq \label{eq:H-QM}
H_\text{int}^\text{QM} = \frac{\lambda \tilde{\Lambda}}{\pi^2}  \left[ \frac{9}{32} a^\dag a^\dag a a - \frac{g_2}{2 g_0} a^\dag a a a + \frac{g_4}{32 g_0} a a a a + H.c. + \mO\left( \epsilon^2 \right) \right], \eeq
where we denote $a \equiv a_1$ for shortness. Recall that Fourier coefficients $g_2$ and $g_4$, which contain the information about the motion of mirrors (see Eq.~\eqref{eq:Fourier}), are small, $g_2 \sim g_4 \sim \epsilon$, in the limit $\epsilon \to 0$.

Keeping in mind the normal-ordered structure of this Hamiltonian and assuming the vacuum initial state, $| in \rangle = | 0 \rangle$, we straightforwardly obtain the evolved quantum state in the limit $\lambda \to 0$, $t \to +\infty$ and $\epsilon \to 0$:
\beq \label{eq:Psi-QM}
\begin{aligned}
| \Psi(t) \rangle &= \mathcal{T} \exp\left( -i \int_{t_0}^t H_\text{int}(t') dt' \right) | in \rangle \approx \exp\left( -i \tilde{t} H_\text{int}^\text{QM} \right) | in \rangle \\ &= | in \rangle + \frac{1}{216}\frac{g_4^*}{g_0} \left[ \exp\left( - i \frac{27}{4 \pi^2} \lambda \tilde{\Lambda} \tilde{t} \right) - 1 \right] \left( a^\dag \right)^4 | in \rangle + \mO\left( \epsilon^2 \right).
\end{aligned} \eeq
The leading contribution to Eq.~\eqref{eq:Psi-QM} is ensured by multiple scattering of virtual particles (i.e., by powers of the $a^\dag a^\dag a a$ term). Note that the $\mathcal{T}$-ordered exponential is resolved into an ordinary exponential because in the limit in question, time-varying parts of the effective Hamiltonian are suppressed by powers of $\lambda$, so the identity $\left[ H(t_1), H(t_2) \right] \approx 0$ approximately holds for large evolution times. In other words, $\mathcal{T}$-exponential is resolved because we want to keep only leading secularly growing terms in its expansion.

Identity~\eqref{eq:Psi-QM} readily implies the approximate level density and anomalous quantum average:
\begin{align}
\label{eq:n-QM}
n(t) &= \langle \Psi(t) | a^\dag a | \Psi(t) \rangle = \frac{2}{243} \frac{|g_4|^2}{g_0^2} \sin^2\left( \frac{27}{8 \pi^2} \lambda \tilde{\Lambda} \tilde{t} \right) + \mO\left( \epsilon^4 \right), \\
\label{eq:k-QM}
\kappa(t) &= \langle \Psi(t) | a a | \Psi(t) \rangle = - \frac{8}{405} \frac{g_2 g_4^*}{g_0^2} \left[ 5 - 6 \exp\left(- i \frac{9}{8 \pi^2} \lambda \tilde{\Lambda} \tilde{t} \right) + \exp\left(- i \frac{27}{4 \pi^2} \lambda \tilde{\Lambda} \tilde{t} \right) \right]  + \mO\left( \epsilon^3 \right).
\end{align}
Substituting~\eqref{eq:n-QM}, \eqref{eq:k-QM}, and~\eqref{eq:B} into Eq.~\eqref{eq:N} and truncating all summations at $n=1$, we find the total number of particles created in the fundamental mode:
\beq \label{eq:N-QM}
\mN_1 = \frac{|g_2|^2}{4 g_0^2} + \frac{2}{243} \frac{|g_4|^2}{g_0^2} \sin^2\left( \frac{27}{8 \pi^2} \lambda \tilde{\Lambda} \tilde{t} \right) + \mO\left( \epsilon^3 \right). \eeq
The first and second terms on the right-hand side of~\eqref{eq:N-QM} describe the tree-level and loop-level contributions, respectively. Note that the loop contribution is always positive and has the same order in $\epsilon$ as the tree-level approximation. Moreover, at very large times, $\tilde{t} \gg 1 / \lambda \tilde{\Lambda}$, the oscillating part can be replaced with the average value:
\beq \mN_1 \approx \frac{|g_2|^2}{4 g_0^2} + \frac{1}{243} \frac{|g_4|^2}{g_0^2}. \eeq
Therefore, nonlinearities enhance the number of particles created in the fundamental mode and measured at large evolution times. Furthermore, the loop contribution significantly surpasses the tree-level approximation in the case $|g_4| \gg |g_2|$, e.g., in resonant oscillations~\eqref{eq:resonant} with $q = 4$.

We emphasize that the loop corrections to the number of created particles are related to a change in the initial quantum state, which becomes possible due to the violation of the energy conservation law in a background field. The leading correction, equation~\eqref{eq:Psi-QM}, describes a state with four correlated particles. Note that the full series in~\eqref{eq:Psi-QM} is obtained by a unitary evolution from a pure state, although unitarity is violated in any fixed order in $\epsilon$.

\section{Large~\texorpdfstring{$N$}{N} generalization}
\label{sec:large-N}

Now let us show that the fundamental mode is indeed the most populous one. To that end, consider the $O(N)$-symmetric, large~$N$ generalization of the model~\eqref{eq:DCE-nlin}:
\beq \label{eq:DCE-large-N}
\left( \pd_t^2 - \pd_x^2 \right) \phi^i = \frac{\lambda}{N} (\phi^j \phi^j) \phi^i, \quad \phi^i[t, L(t)] = \phi^i[t, R(t)] = 0, \eeq
where $i = 1, \cdots\!, N$ with $N \gg 1$, and we assume the summation over the repeated upper indices. The effective Hamiltonian of this model in the limit $\lambda \to 0$, $t \to \infty$ and $\epsilon \to 0$ coincides with the Hamiltonian~\eqref{eq:H} after the appropriate change in the operator products:
\beq \begin{gathered}
3 a_k^\dag a_l^\dag a_m a_n \to (a_k^i)^\dag (a_l^i)^\dag a_m^j a_n^j + (a_k^i)^\dag (a_l^j)^\dag a_m^i a_n^j + (a_k^i)^\dag (a_l^j)^\dag a_m^j a_n^i, \\
a_k^\dag a_l a_m a_n \to (a_k^i)^\dag a_l^i a_m^j a_n^j, \qquad
a_k a_l a_m a_n \to a_k^i a_l^i a_m^j a_n^j, \\
3 a_k^\dag a_l \to (N+2) (a_k^i)^\dag a_l^i, \qquad 3 a_k a_l \to (N+2) a_k^i a_l^i,
\end{gathered} \eeq
and coupling constant, $\lambda \to \lambda / N$. Similarly to the original Hamiltonian~\eqref{eq:H}, the main contribution to the generalized $O(N)$-symmetric Hamiltonian is ensured by elastic scattering processes\footnote{In the Hamiltonian~\eqref{eq:H-0}, particle decays are suppressed by a numerical factor of the order of 5. The large~$N$ limit of the $O(N)$-symmetric generalization simply enhances this factor to $5N$.}, and the leading relevant correction to this Hamiltonian is given by the production of four correlated quadruplets. Due to this reason, we believe that the qualitative behavior of both models are approximately the same, at least at small deviations from stationarity.

At the same time, the evolved quantum state in the model~\eqref{eq:DCE-large-N} is straightforwardly calculated:
\beq \label{eq:Psi-large-N}
\begin{aligned} | \Psi(t) \rangle \approx | in \rangle &- \frac{1}{N} \sum_{p,q} \frac{g_{p+q}^*}{g_0} \frac{A_{p,q}(t)}{\sqrt{pq}} (a_p^i)^\dag (a_q^i)^\dag | in \rangle \\ &+ \frac{1}{N} \sum_{k,l,m,n} \frac{g_{k+l+m+n}^*}{4 g_0} \frac{B_{k,l,m,n}(t)}{\sqrt{klmn}} (a_k^i)^\dag (a_l^i)^\dag (a_m^j)^\dag (a_n^j)^\dag | in \rangle + \mO\left( \epsilon^2\right) + \mO\left( \frac{1}{N^2} \right), \end{aligned} \eeq
where
\begin{align}
&\begin{aligned}
A_{p,q}(t) = &\sum_{k,l,m,n} \frac{\delta_{p+q,k+l+m+n}}{klm} \bigg[ \delta_{q,n} \frac{e^{-i \tau C_{k,l,m,n}} + i \tau C_{k,l,m,n} - 1}{C_{k,l,m,n}^2} \\ &+ \frac{1}{n(k+l+m)} \left( \frac{e^{-i \tau D_{p,q}} - 1}{D_{p,q}^2 \left( D_{p,q} - C_{k,l,m,n} \right)} - \frac{e^{-i \tau C_{k,l,m,n}} - 1}{C_{k,l,m,n}^2 \left( D_{p,q} - C_{k,l,m,n} \right)} - \frac{i \tau}{D_{p,q} C_{k,l,m,n}} \right) \bigg],
\end{aligned} \\
&B_{k,l,m,n}(t) = \frac{e^{ - i \tau C_{k,l,m,n}} - 1}{2 C_{k,l,m,n}}.
\end{align}
For shortness, we rescale the time, $\tau = \frac{\lambda \tilde{\Lambda} \tilde{t}}{4 \pi^2}$, introduce coefficients $C_{k,l,m,n} = D_{k,l} + D_{m,n}$ and $D_{p,q} = \frac{H_{p+q-1}}{p+q} + \frac{1}{pq}$, and denote the Harmonic numbers as $H_n = \sum_{k=1}^n \frac{1}{k} \sim \log n + \gamma$, where $\gamma$ is the Euler-Mascheroni constant. Note that the contribution of the states with more than four particles is suppressed by the powers of $1/N$.

Substituting the final state~\eqref{eq:Psi-large-N} into Eqs.~\eqref{eq:n} and~\eqref{eq:k}, we obtain the future asymptotics of the resummed quantum averages:
\begin{gather}
\label{eq:n-large-N}
n_{pq}^{ij}(t) \approx \frac{\delta_{i,j}}{N} \frac{2}{\sqrt{pq}} \sum_{k,l,m} \frac{g_{p,k,l,m}^* g_{q,k,l,m}}{g_0^2 \, klm} B_{p,k,l,m}^*(t) B_{q,k,l,m}(t) + \mO\left( \epsilon^3 \right) + \mO \left( \frac{1}{N^2} \right), \\ 
\label{eq:k-large-N}
\kappa_{pq}^{ij}(t) \approx -\frac{\delta_{i,j}}{N} \frac{1}{\sqrt{pq}} \frac{g_{p+q}^*}{g_0} \Big[ A_{p,q}(t) + A_{q,p}(t) \Big] + \mO\left( \epsilon^2 \right) + \mO \left( \frac{1}{N^2} \right),
\end{gather}
and the number of created particles:
\beq \label{eq:N-largeN-gen}
\begin{aligned}
\mN_n(t) \approx N \sum_k \frac{|g_{n+k}|^2}{g_0^2} \frac{n k}{(n+k)^2} &+ 2 \sum_{k,l,m} \frac{|g_{n+k+l+m}|^2}{g_0^2} \frac{B_{n,k,l,m}^*(t) B_{n,k,l,m}(t)}{n k l m} \\ &+ 2 \sum_{k} \frac{|g_{n+k}|^2}{g_0^2} \frac{\text{Re}\left[ A_{n,k}(t) + A_{k,n}(t) \right]}{n+k} + \mO \left( \epsilon^3 \right) + \mO \left( \frac{1}{N^2} \right).
\end{aligned} \eeq
The first term in Eq.~\eqref{eq:N-largeN-gen} describes the tree-level contribution, $\mN_n^\text{free}$; the second and third terms appear only in the nonlinear, interacting theory.

\begin{figure}[t]
    \center{\includegraphics[scale=0.75]{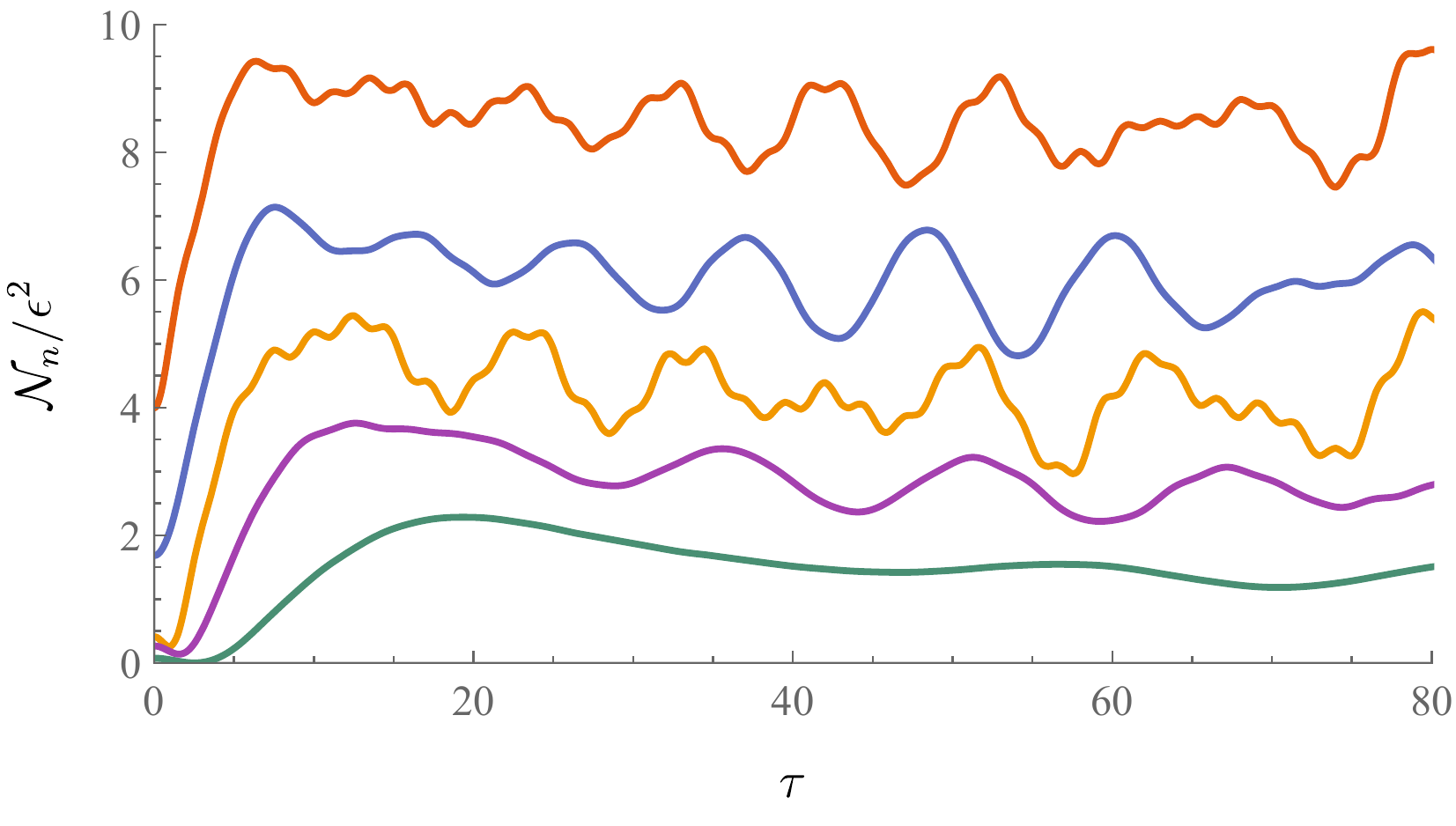}}
    \caption{Numerically calculated function $\mN_n(\tau)/\epsilon^2$, $n=1,2,4,8,16$ (from top to bottom) for the resonant motion~\eqref{eq:resonant} with $q=2$ and $\epsilon = 0.02$. The ultraviolet cutoff is set to $n_\text{UV} = 64$. For illustrative purposes, we also set $N = 1$; in this case, the ``initial'' values of functions coincide with the normalized tree-level contribution, $\mN_n(0) = \mN_n^\text{free} / N$.}
    \label{fig:numerics}
\end{figure}

Let us also explicitly evaluate $\mN_n(t)$ for a particular physically meaningful mirror motion. As an example of such a motion, we consider resonant oscillations with $q = 2$ frequency and destructive interference~\eqref{eq:resonant}. We approximately calculate the particle number for relatively small, $\lambda \tilde{\Lambda} \tilde{t} \ll 1$, and large, $\lambda \tilde{\Lambda} \tilde{t} \gg n_\text{UV}/\log( n_\text{UV} )$, evolution times, where $n_\text{UV}$ defines the effective ultraviolet cutoff. At intermediate times, these asymptotics are connected by a smooth curve, which can be calculated numerically (Fig.~\ref{fig:numerics}).

\begin{figure}[t]
\begin{minipage}[h]{0.49\linewidth}
\center{\includegraphics[width=1\linewidth]{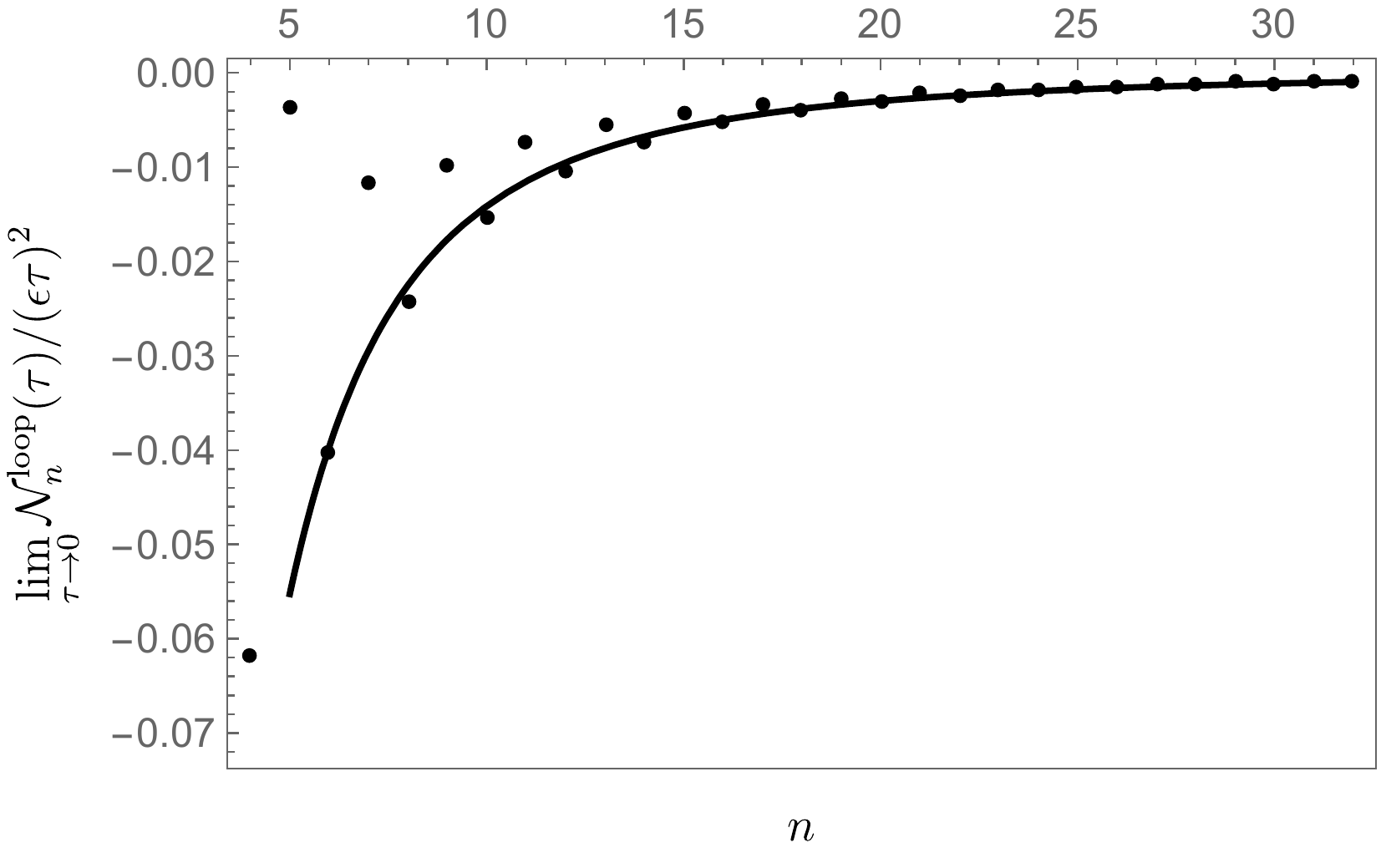} \\ (a)}
\end{minipage}
\hfill
\begin{minipage}[h]{0.49\linewidth}
\center{\includegraphics[width=1\linewidth]{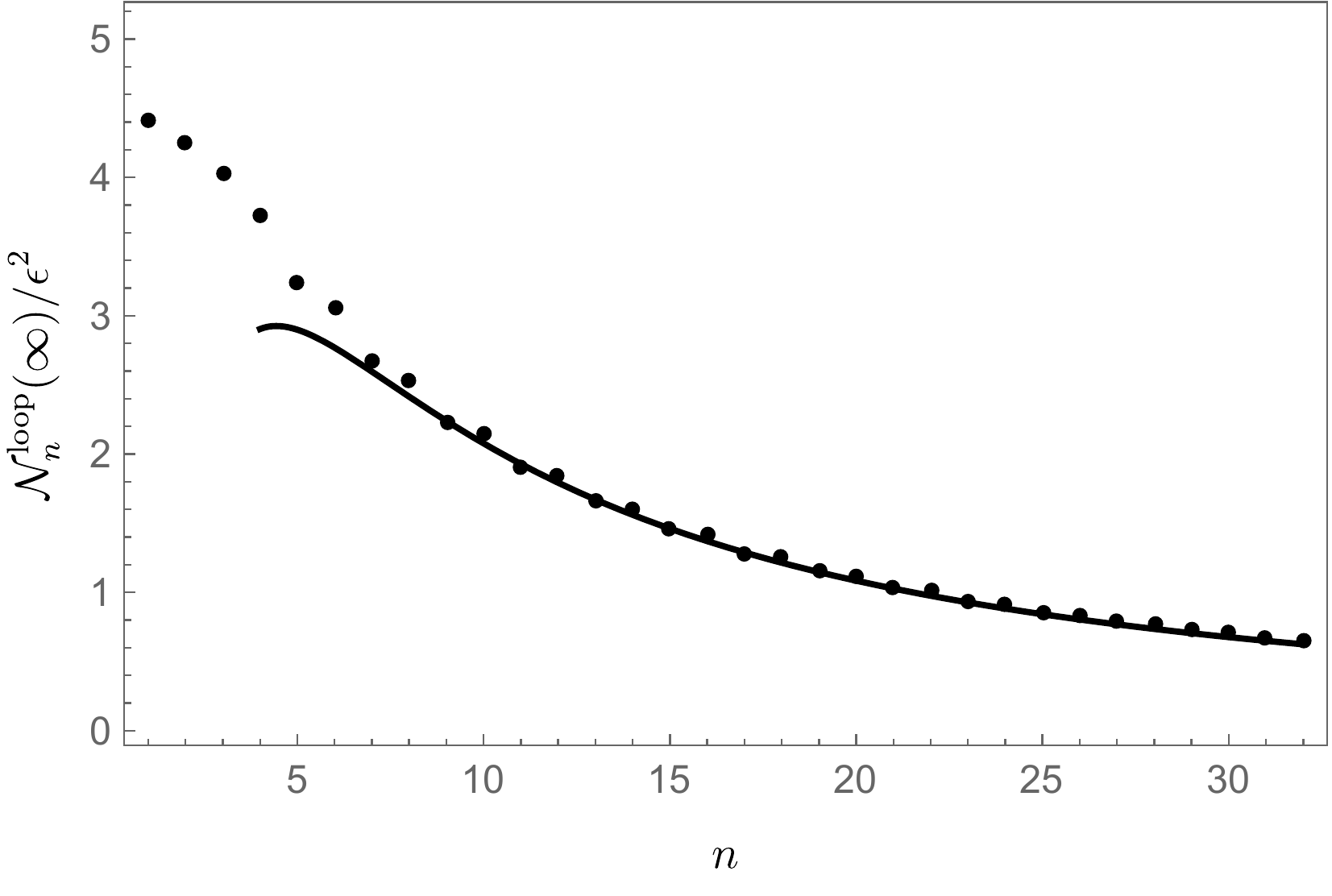} \\ (b)}
\end{minipage}
\caption{Numerically estimated functions $\mN_n^\text{loop}(\tau)/(\epsilon \tau)^2$, as $\tau \to 0$ (a), and $\mN_n^\text{loop}(\tau)/\epsilon^2$, as $\tau \to \infty$ (b), for the resonant motion~\eqref{eq:resonant} with $q = 2$ and $\epsilon = 0.02$. The ultraviolet cutoff is set to $n_\text{UV} = 64$. The dots denote the calculated values, the lines denote the fitting curves. Note that the first graph starts from $n=4$; the values at $n < 4$ are positive and much larger than the values at $n \ge 4$.}
\label{fig:fit}
\end{figure}

At relatively small evolution times, $\lambda \tilde{\Lambda} \tilde{t} \ll 1$, the oscillating functions in~\eqref{eq:N-largeN-gen} can be expanded in a series. The next-to-the-leading term in this series, which originates from the two-loop correction to the quantum averages, quadratically grows with time (compare with~\cite{Alexeev, Akopyan}):
\beq \label{eq:N-largeN-small}
\begin{aligned}
\mN_n &\approx \mN_n^\text{free} + \frac{\tau^2}{2} \sum_{k,l,m} \frac{|g_{n+k+l+m}|^2}{g_0^2} \frac{1}{n k l m} - \tau^2 \sum_{k,l,m,p} \frac{|g_{n+k}|^2}{g_0^2} \frac{\delta_{n,l+m+p} + \delta_{k,l+m+p}}{(n+k) l m p} \\
&\sim \frac{\epsilon^2}{n} \left[ N - 2 \tau^2 \frac{\log^2 n}{n^2} + \cdots \right], \quad \text{for} \quad 1 \ll n \ll \frac{1}{\epsilon}.
\end{aligned} \eeq
In the last line, we discard the overall factor and fit the loop contribution, $\mN_n^\text{loop} = \mN_n - \mN_n^\text{free}$, from the numerically calculated dependence (Fig.~\ref{fig:fit}a). We emphasize that the total number of created particles is always positive, although the loop contribution is negative for $n \ge 4$ and small evolution times (see Fig.~\ref{fig:numerics}).

At large evolution times, $\lambda \tilde{\Lambda} \tilde{t} \gg n_\text{UV} / \log( n_\text{UV} )$, we can replace the oscillating functions with their average values:
\beq \label{eq:N-largeN-large}
\begin{aligned}
\mN_n &\approx \mN_n^\text{free} + \sum_{k,l,m} \frac{|g_{n+k+l+m}|^2}{g_0^2} \frac{1}{n k l m} \frac{1}{C_{n,k,l,m}^2} \\ &\phantom{\approx \mN_n^\text{free} \;}+ \sum_{k,l,m,p,q} \frac{|g_{n+k}|^2}{g_0^2} \frac{\delta_{n+k,l+m+p+q}}{(n+k) l m p} \left[ \frac{4}{q (l+m+p)} \frac{D_{n,k} + C_{l,m,p,q}}{D_{n,k}^2 C_{l,m,p,q}^2} - \frac{2 \delta_{k,q} + 2 \delta_{n,q}}{C_{l,m,p,q}^2} \right] \\ &\sim \frac{\epsilon^2}{n} \left[ N + C \frac{\log^2 n}{n} + \cdots \right], \quad \text{for} \quad 1 \ll n \ll \frac{1}{\epsilon},
\end{aligned} \eeq
where we again discard the overall factor and fit the loop contribution from the numerically calculated dependence (Fig.~\ref{fig:fit}b). The factor $C$ in front of the loop contribution is positive and slightly changes with the ultraviolet cutoff, $C \sim \log \log(n_\text{UV})$. For reasonable values of $n_\text{UV}$, this factor has the order of $C \sim 100$.

It is worth mentioning that Eqs.~\eqref{eq:N-largeN-small} and~\eqref{eq:N-largeN-large} are valid only for not very high mode numbers, $n \ll 1/\epsilon$, where the Bogoliubov coefficients can be approximated by Eqs.~\eqref{eq:B}. However, we emphasize that for every $n \ll 1/\epsilon$ loop contribution to the created particle number have the same order in $\epsilon$ as the tree-level expression.

We also note that the number of created particles is proportional to a small factor, $\epsilon^2$, and goes to zero at high frequencies, $\mN_n \to 0$ as $n \to \infty$. This qualitatively validates the low-particle approximation of Sec.~\ref{sec:QM}.

\section{Nonlinear DCE in a Josephson metamaterial}
\label{sec:circuit}

Now let us apply the approach of Secs.~\ref{sec:Bogoliubov}--\ref{sec:QM} to another model of the nonlinear DCE, which has a clear experimental implementation~\cite{Lahteenmaki}. Namely, consider the Hamiltonian of a Josepshon metamaterial, i.e., an array of superconducting quantum interference devices (dc-SQUIDs) in a varying external magnetic field:
\beq \label{eq:HC}
H_{JM} \approx \int_0^\Lambda \left[ \frac{1}{2} (\pd_t \phi)^2 + \frac{v^2(t,x)}{2} (\pd_x \phi)^2 - \frac{\lambda v^2(t,x)}{4} (\pd_x \phi)^4 \right], \quad \phi(t,0) = \phi(t,\Lambda) = 0. \eeq
The effective cavity length $\Lambda$, speed of light $v(t,x)$, and coupling constant $\lambda$ are specified by the parameters of the metamaterial (see Appendix~\ref{sec:Josephson} for the details). Moreover, variations of the external magnetic field directly translate into the variations of the light speed $v(t,x)$. We assume that these variations are small, nonresonant, and vanish at the past and future infinity, i.e., $v(t,x) = v_\infty + \epsilon \tilde{v}(t,x)$, where $v_\infty = \const$, $\epsilon \ll 1$ and $\tilde{v}(t,x) \to 0$ as $t \to \pm \infty$. In other words, we consider the Hamiltonian~\eqref{eq:HC} for small deviations from stationarity.

First of all, we split the Hamiltonian~\eqref{eq:HC} into the free (quadratic) and interacting (quartic) parts and quantize the free model similarly to Sec.~\ref{sec:Bogoliubov}:
\beq \phi(t,x) = \sum_{n=1}^{n_\text{UV}} \left[ a_n f_n^\text{in}(t,x) + H.c. \right]. \eeq
Here, $a_n^\dagger$ and $a_n$ are the standard bosonic creation and annihilation operators; in-modes $f_n^\text{in}(t,x)$ solve the corresponding free equations of motion, form the complete orthonormal basis with respect to the Klein-Gordon inner product and have definite positive frequency at the past infinity\footnote{Note that in this section, we explicitly keep the Planck constant $\hbar$ and the ultraviolet cutoff $n_\text{UV} \sim N$, where $N$ is the total number of nodes in the circuit.}:
\beq f_n^\text{in}(t,x) \to \sqrt{\frac{\hbar}{\pi n v_\infty}} \exp\left( -i \pi n \frac{v_\infty t}{\Lambda} \right) \sin\left( \frac{\pi n x}{\Lambda} \right), \quad \text{as} \quad t \to -\infty. \eeq
At the future infinity, in-modes are rewritten in terms of the out-modes:
\beq f_n^\text{in}(t,x) \to \sum_{k=1}^{n_\text{UV}} \sqrt{\frac{\hbar}{\pi k v_\infty}} \left[ \alpha_{nk} \exp\left( -i \pi k \frac{v_\infty t}{\Lambda} \right) + \beta_{nk} \exp\left( i \pi k \frac{v_\infty t}{\Lambda} \right) \right] \sin\left( \frac{\pi k x}{\Lambda} \right), \quad \text{as} \quad t \to +\infty, \eeq
where Bogoliubov coefficients $\alpha_{nk}$ and $\beta_{nk}$ are determined by the variations of the speed of light. If these variations are small and nonresonant, Bogoliubov coefficients are close to the identity, $\alpha_{nk} \approx \delta_{n,k} + \epsilon \tilde{\alpha}_{nk} \sqrt{n/k}$ and $\beta_{nk} \approx \epsilon \tilde{\beta}_{nk} \sqrt{n/k}$ (compare with~\eqref{eq:B}).

Then we substitute the in-modes into the interacting Hamiltonian, expand it to the first order in $\epsilon$, take the integral over $dx$ and neglect rapidly oscillating terms:
\begin{align}
H_\text{int}^{(0)} &= - \frac{\lambda \pi \hbar^2}{32 \Lambda^3} \sum_{k,l,m,n} \sqrt{k l m n} \Big[ 4 \delta_{k,l+m+n} a_k^\dagger a_l a_m a_n + \left( 3 \delta_{k+l,m+n} + 6 \delta_{k,m} \delta_{l,n} \right) a_k^\dagger a_l^\dagger a_m a_n + H.c. \Big], \\
H_\text{int}^{(1)} &= - \epsilon \frac{\lambda \pi \hbar^2}{32 \Lambda^3} \sum_{k,l,m,n} \sqrt{k l m n} \Big[ 4 \tilde{\beta}_{k,l+m+n} a_k a_l a_m a_n + f_{k,l,m,n}^{(1,3)} a_k^\dagger a_l a_m a_n + f_{k,l,m,n}^{(2,2)} a_k^\dagger a_l^\dagger a_m a_n + H.c. \Big],
\end{align}
as $t \to +\infty$. Here, $f_{k,l,m,n}^{(1,3)}$ and $f_{k,l,m,n}^{(2,2)}$ are some constant $\mO(1)$ tensors that depend on the Bogoliubov coefficients; we do not need their explicit form to determine the leading correction to the number of created quasiparticles in the limit $\epsilon \ll 1$. In fact, in this limit, the leading loop contributions to the level density~\eqref{eq:n} and anomalous quantum average~\eqref{eq:k} are approximated by the following expressions (we set $t_0 = 0$ for shortness):
\begin{align}
n_{kl}(t) &\approx \Big\langle in \Big| \sum_{a=1}^\infty \frac{1}{a!} \left( \frac{i t}{\hbar} \right)^a H_\text{int}^{(1)} \left( H_\text{int}^{(0)} \right)^{a-1} a_k^\dagger a_l \sum_{b=1}^\infty \frac{1}{b!} \left( - \frac{i t}{\hbar} \right)^b \left( H_\text{int}^{(0)} \right)^{b-1} H_\text{int}^{(1)} \Big| in \Big\rangle, \\
\kappa_{kl}(t) &\approx \Big\langle in \Big| \sum_{a=0}^\infty \frac{1}{a!} \left( \frac{i t}{\hbar} \right)^a \left( H_\text{int}^{(0)} \right)^a a_k a_l \sum_{b=1}^\infty \frac{1}{b!} \left( - \frac{i t}{\hbar} \right)^b \left( H_\text{int}^{(0)} \right)^{b-1} H_\text{int}^{(1)} \Big| in \Big\rangle.
\end{align}
Keeping in mind that at small deviations from the stationarity, created quasiparticles mainly populate the lowest-energy mode (compare with Secs.~\ref{sec:QM} and~\ref{sec:large-N}), we truncate the summations at $n_\text{UV} = 1$ for a rough estimate of the effective Hamiltonian:
\beq H_\text{int} \approx - \frac{\lambda \pi \hbar^2}{32 \Lambda^3} \left[ 9 a^\dagger a^\dagger a a + 4 \epsilon \tilde{\beta}_{1,3} a a a a + \epsilon f_{1,1,1,1}^{(1,3)} a^\dagger a a a + \epsilon f_{1,1,1,1}^{(2,2)} a^\dagger a^\dagger a a + H.c. + \mO(\epsilon^2) \right]. \eeq
This Hamiltonian qualitatively coincides with the Hamiltonian~\eqref{eq:H-QM}. The leading order, $\mO(1)$, term describes the elastic scattering of quasiparticles; the subleading, $\mO(\epsilon)$, terms describe the processes that violate the energy conservation law, i.e., borrow the energy of the external magnetic field to create new quasiparticles. The most important process is the simultaneous production of four correlated excitations, which is determined by the Bogoliubov coefficient $\tilde{\beta}_{1,3}$ and provides the leading correction to the number of created quasiparticles:
\beq \label{eq:JM-loop}
\mN_1^\text{loop} = \frac{32}{243} \epsilon^2 \big| \tilde{\beta}_{1,3} \big|^2 \sin^2 \left( \frac{27}{8} \frac{\lambda \pi \hbar t}{\Lambda^3} \right) + \mO\left(\epsilon^4\right). \eeq
This establishes a qualitative equivalence of the moving mirror, Eq.~\eqref{eq:DCE-nlin}, and Josepshon metamaterial, Eq.~\eqref{eq:DCE-squid}, models of the DCE (at least for small deviations from the stationarity).

We emphasize that at large evolution times, $t \gg t_* \sim \Lambda^3/\lambda \hbar$, the loop contribution is comparable to the tree-level value, $\mN_1^\text{free} \approx \epsilon^2 \big| \tilde{\beta}_{1,1} \big|^2$. For realistic parameters of the Josephson metamaterial\footnote{\label{footnote} In the experiment~\cite{Lahteenmaki}, authors created a Josephson metamaterial with the following parameters: length $\Lambda = 4$ mm, number of nodes $N = 250$, Josephson critical current $I_\text{c} \sim 10^{-5}$ A, and the effective speed of light in the unperturbed circuit $v_\infty \sim 0.5 c$, with $c$ being the speed of light in the vacuum.}, this time is approximately $t_* \sim 10^{-5}$ s. In other words, small intrinsic nonlinearities of the dc-SCQUIDs are negligible only at relatively small times, $t \ll t_*$; at larger times, the measured number of created particles will deviate from the tree-level value\footnote{We remind that we resum only the leading secularly growing corrections in the limit $\lambda \to 0$, $t \to \infty$, $\lambda t = \const$, i.e., keep only the terms of the form $(\lambda t)^k$ with an integer $k > 0$. The subleading corrections has the form $\lambda^m t^n$ with some integers $m > n > 0$; at finite $\lambda$ and $t$, these contributions are suppressed by the powers of $\lambda$. For realistic parameters of the model, see footnote~\ref{footnote} and paper~\cite{Lahteenmaki}, these subleading terms become significant only at the time scale $t_{**} \sim \Lambda^5 v_\infty / \lambda^2 \hbar^2 \sim 1$ s, which is much larger that $t_*$.}.

\section{Discussion and Conclusion}
\label{sec:discussion}

In this work, we have shown that nonlinearities (i.e., interactions between the quantum fields) significantly affect particle creation in a weakly nonstationary scalar DCE. We emphasize that at large evolution times, the correction to the particle number is comparable to the tree-level value even if the coupling constant is small, i.e., nonlinearities are apparently weak. First of all, we calculated the Bogoliubov coefficients and derived the effective Hamiltonian of the quantized field. Then we employed this Hamiltonian to evaluate the nonperturbative (in the coupling constant~$\lambda$) correction to the tree-level number of created particles. We also discussed the effective Hamiltonian of a Josephson metamaterial, where DCE can be measured experimentally. In both cases, created particles mainly populate the fundamental (lowest-energy) mode. More importantly, in both cases tree-level and loop-level contributions to the number of created particles have the same order in the small parameter $\epsilon$, which determines the magnitude of deviations from stationarity.

We emphasize that in this paper, we work in the interaction picture. In other words, we separate the free, Eq.~\eqref{eq:H-free}, and interacting, Eq.~\eqref{eq:H}, Hamiltonians and assume that we \textit{exactly} know how the free Hamiltonian evolves the field operator $\phi(t)$. This knowledge is contained in the Bogoliubov coefficients. Then we approximately calculate the final quantum state, i.e., evolve the initial quantum state with the interacting Hamiltonian. This information is contained in the resummed level density and anomalous quantum average. Note that the quantum state continues evolving even when the Bogoliubov coefficients stop changing, i.e., free evolution is essentially finished. This approach is inspired by similar problems from the high energy physics~\cite{Akhmedov:dS, Krotov, Polyakov, Bascone, Pavlenko, Moschella, Bazarov, Musaev, Akhmedov:Et, Akhmedov:Ex, Diatlyk-1, Lanina, Diatlyk-2, Akhmedov:H, Trunin-1, Trunin-QM, Akhmedov:2021}.

Note that we restricted ourselves to the calculation of the number of created particles, Eq.~\eqref{eq:N} with the resummed level density~\eqref{eq:n} and anomalous quantum average~\eqref{eq:k}. Nevertheless, our analysis can be straightforwardly applied to the calculation of the resummed Keldysh propagator, which is also expressed through the quantum averages~\eqref{eq:n} and~\eqref{eq:k}. Using this propagator, one can easily obtain other observable quantities (e.g., see Appendix~\ref{sec:N-phys}).

Besides, we emphasize that the summation of loop corrections in Secs.~\ref{sec:QM}--\ref{sec:circuit} corresponds to the solution of a reduced system of Schwinger-Dyson equations on exact correlation functions written in the Schwinger-Keldysh technique. Namely, for this summation, we single out only ``tadpole'' and ``bubble'' contributions to propagators and vertices, respectively. In Secs.~\ref{sec:QM} and~\ref{sec:circuit}, such a reduction is achieved by employing the limit $\epsilon \to 0$ and $\lambda \to 0$, $t \to \infty$; in the model of Sec.~\ref{sec:large-N}, it naturally emerges in the large~$N$ limit. Then, as soon as the exact Keldysh propagator is calculated, quantum averages and created particle number are unambiguously restored. More details on the relation between the different approaches to the summation of loop corrections to the created particle number can be found in~\cite{Trunin-QM}, where a simplified model of the DCE with a similar diagrammatics is discussed.

Our work can be extended in several possible directions. First, we expect that the effects studied in this paper can be measured in a quantum circuit analog of the DCE~\cite{Wilson, Lahteenmaki, Nation} (see also~\cite{Johansson:2010, Nation:2008, Leib, Nigg, Bourassa, Krupko, Blais}). Namely, we expect that at large evolution times, $t \gg t_* \sim 10^{-5}$ s, the number of the quasiparticles created in a quantum circuit will deviate from the tree-level value due to the intrinsic nonlinearity of Josephson junctions (see Sec.~\ref{sec:circuit}). In other words, this means that the Gaussian approximation used in~\cite{Wilson, Lahteenmaki, Nation} for a theoretical estimate of $\mN$ is valid only at relatively small times.

However, we remind that in this paper, we considered only small deviations from the stationarity. At the same time, feasible experimental implementations of models~\eqref{eq:DCE-nlin} or~\eqref{eq:DCE-squid} require a resonant pumping of energy into the system, i.e., strongly deviate from the stationarity. In this case, Bogoliubov coefficients and number of created particles rapidly grow with time already at the tree level~\cite{Dodonov:1996, Dodonov:2010, Dodonov:2020, Nation, Dalvit:1997, Dalvit:1998, Dodonov:1998}. Therefore, it is promising to extend the results of the present paper to strong deviations from stationarity, including the resonant pumping of energy into the system. In the large~$N$ limit, such an extension can be possibly implemented using the Schwinger-Keldysh technique similarly to papers~\cite{Akhmedov:dS, Akhmedov:Et, Akhmedov:Ex, Trunin-QM}. We expect that in strongly nonstationary systems, nonlinearities also significantly affect particle production through the generation of quantum averages $n_{kl}$ and $\kappa_{kl}$ because these averages are eventually multiplied by large Bogoliubov coefficients and amplified on a par with vacuum fluctuations, see Eq.~\eqref{eq:N} (also compare with the simplified model of~\cite{Trunin-QM} where calculations for strong deviations from stationarity were performed).

Second, the expression for the effective Hamiltonian is easily generalized to other realistic models of the DCE, including massive quantum fields, semitransparent mirrors, and resonant oscillations with constructive interference. In general, the strategy is similar to Secs.~\ref{sec:Bogoliubov} and~\ref{sec:H}: one calculates the Bogoliubov coefficients for the free field and substitutes them into the Hamiltonian~\eqref{eq:H-app}.

Third, our analysis can be extended to other nonequilibrium quantum systems with a ``nonkinetic'' behavior of loop corrections, such as light self-interacting scalar fields in the de Sitter space~\cite{Pavlenko, Moschella, Serreau-1, Serreau-2, Serreau-3, Nacir} or upper Rindler wedge~\cite{Bazarov}.

Finally, two-dimensional $O(N)$ and $\mathbb{C}P^N$ models on a finite interval are known to possess nontrivial classical solutions with a negative energy~\cite{Gorsky, Pikalov, Milekhin-1, Milekhin-2, Bolognesi, Nitta, Flachi-1}. In these models, the naive vacuum, $a_n | in \rangle = 0$, is unstable at some values of the model parameters. In principle, nonequilibrium techniques (similar to those used in Sec.~\ref{sec:large-N}) might confirm this instability and explicitly trace the evolution from the initial quantum state to the true vacuum. An example of such a calculation in a different setup can be found in~\cite{Diatlyk-2}. Furthermore, it is promising to generalize the predictions of~\cite{Gorsky, Pikalov, Milekhin-1, Milekhin-2, Bolognesi, Nitta, Flachi-1, Flachi-2}, which were obtained in a stationary approximation, to time-dependent boundary conditions. The calculations of Sec.~\ref{sec:large-N} may be considered as a starting point for such generalizations.

\section*{Acknowledgments}

I would like to thank Emil Akhmedov for the fruitful discussions and sharing his ideas, Alexander Gorsky for the useful comments, and Elena Bazanova for the proofreading of the text. This work was supported by the Russian Ministry of Education and Science and by the grant from the Foundation for the Advancement of Theoretical Physics and Mathematics ``BASIS''.

\appendix

\section{Alternative derivation of the created particle number}
\label{sec:N-phys}

Correlation functions are the most fundamental objects in the nonstationary quantum field theory~\cite{Akhmedov:2021}. First, they have a clear physical meaning even in time-dependent backgrounds where the notion of a particle is ill-defined. Second, all directly observable quantities are derived from correlation functions. For example, the exact Hamiltonian of the nonlinear DCE~\eqref{eq:DCE-nlin} can be obtained from the exact Keldysh propagator:
\beq H_\text{full}(t) = \int_{L(t)}^{R(t)} dx_1 \left[ \frac{1}{2} \pd_{t_1} \pd_{t_2} G^K(\mathbf{x}_1, \mathbf{x}_2) + \frac{1}{2} \pd_{x_1} \pd_{x_2} G^K(\mathbf{x}_1, \mathbf{x}_2) + \frac{\lambda}{4} \left( G^K(\mathbf{x}_1, \mathbf{x}_2) \right)^2 \right]_{\mathbf{x}_1 \to \mathbf{x}_2}, \eeq
where we denote $\mathbf{x} = (t,x)$ for shortness. At the tree level, the Keldysh propagator is defined as an anticommutator of quantum fields:
\beq \label{eq:GK}
\begin{aligned}
G_\text{free}^K(\mathbf{x}_1, \mathbf{x}_2) &= \frac{1}{2} \big\langle in \big| \left\{ \phi(\mathbf{x}_1), \phi(\mathbf{x}_2) \right\} \big| in \big\rangle \\ &= \sum_{k,l=1}^\infty \left[ \left( \frac{1}{2} \delta_{k,l} + n_{kl} \right) f_k^\text{in}(\mathbf{x}_1) \left(f_l^\text{in}(\mathbf{x}_2)\right)^* + \kappa_{kl} \, f_k^\text{in}(\mathbf{x}_1) f_l^\text{in}(\mathbf{x}_1) + H.c. \right], 
\end{aligned} \eeq
where we employ the mode decomposition~\eqref{eq:phi} and use short notations for the level density and anomalous quantum average. In the interacting theory, the Keldysh propagator preserves the form~\eqref{eq:GK} with exact quantum averages~\eqref{eq:n} and~\eqref{eq:k} if the difference between the times is much smaller than their average, $t_1 - t_2 \ll (t_1 + t_2)/2$. This allows us to estimate the future asymptotics of the exact Hamiltonian in the interacting theory:
\beq H_\text{full} \approx \sum_{k,l,m=1}^\infty \frac{\pi m}{2 \tilde{\Lambda}} \left[ \left( \frac{1}{2} \delta_{k,l} + n_{kl} \right) \left( \alpha_{km} \alpha_{lm}^* + \beta_{km} \beta_{lm}^* \right) + \kappa_{kl} \left( \alpha_{km} \beta_{lm} + \beta_{km} \alpha_{lm} \right) + H.c. \right] + \mO\left( \lambda \tilde{\Lambda} \right), \eeq
where we take the limit $t \to +\infty$, substitute the asymptotics of the in-modes~\eqref{eq:in-future}, rotate to the $(\tilde{t}, \tilde{x})$ coordinates and integrate over $\tilde{x}$. Note that the contribution of the nonlinear part is negligible if we consider small coupling constants, $\lambda \ll 1/\tilde{\Lambda}^2$.

Finally, using the properties of the Bogoliubov coefficients:
\beq \sum_k \left( \alpha_{ik} \alpha_{jk}^* - \beta_{ik} \beta_{jk}^* \right) = \delta_{i,j}, \qquad \sum_k \left( \alpha_{ik} \beta_{jk} - \beta_{ik} \alpha_{jk} \right) = 0 \eeq
and symmetry of the level density, $n_{kl}^* = n_{lk}$, and anomalous quantum average, $\kappa_{kl} = \kappa_{lk}$, we obtain the following expression for the exact Hamiltonian:
\beq \label{eq:H-full}
H_\text{full} \approx \sum_{m=1}^\infty \frac{\pi m}{\tilde{\Lambda}} \left( \frac{1}{2} + \mN_m \right) + \mO\left( \lambda \tilde{\Lambda} \right), \eeq
where $\mN_m$ is given by equation~\eqref{eq:N} with level density~\eqref{eq:n} and anomalous quantum average~\eqref{eq:k}. The divergent sum $\sum_{m=1}^\infty \frac{\pi m}{2 \tilde{\Lambda}} $ describes the standard static Casimir energy and does not depend on the mirrors motion at intermediate times. The second sum describes excitations above stationarity. This confirms that $\mN_m$ has the meaning of the particle number created in a linear or nonlinear DCE at large evolution times.

\section{Full expression for the interacting Hamiltonian}
\label{sec:H-full}

The full expression for the interacting Hamiltonian of theory~\eqref{eq:DCE-nlin} in the limit  $\lambda \to 0$, $t \to +\infty$ has the following form:
\beq \label{eq:H-app}
H_\text{int} = \frac{\lambda \tilde{\Lambda}}{32 \pi^2} \sum_{i,j,k,l,m,n,p,q=1}^\infty \frac{1}{\sqrt{ijkl}} A_{m,n,p,q}^{i,j,k,l}, \eeq
where
\begin{align*}
A_{m,n,p,q}^{i,j,jk,l} &= \left( 3 a_m^\dag a_n^\dag a_p a_q + 6 a_{(n}^\dag \delta_{m),(p}^{\phantom{\dag}} a_{q)}^{\phantom{\dag}} \right) \Big[ \alpha^*_{mi} \alpha^*_{nj} \alpha_{pk} \alpha_{ql} \left( \delta_{i+j,k+l} + \delta_{i,k} \delta_{j,l} + \delta_{i,l} \delta_{j,k} \right) \\ &\quad- 4 \alpha^*_{mi} \alpha^*_{nj} \alpha_{pk} \beta_{ql} \, \delta_{i+j+l,k} + 2 \alpha^*_{mi} \beta^*_{nj} \alpha_{pk} \beta_{ql} \left( \delta_{i+l,j+k} + \delta_{i,j} \delta_{k,l} + \delta_{i,k} \delta_{j,l} \right) \\ &\quad+ 2 \alpha^*_{mi} \beta^*_{nj} \alpha_{pk} \beta_{ql} \left( \delta_{i+k,j+l} + \delta_{i,j} \delta_{k,l} + \delta_{i,l} \delta_{j,k} \right) - 4 \alpha^*_{mi} \beta^*_{nj} \beta_{pk} \beta_{ql} \, \delta_{i+k+l,j} \\ &\quad+ \beta^*_{mi} \beta^*_{nj} \beta_{pk} \beta_{ql} \left( \delta_{i+j,k+l} + \delta_{i,k} \delta_{j,l} + \delta_{i,l} \delta_{j,k} \right) \Big] \\
&- \left(4 a_m^\dag a_n a_p a_q + 6 \delta_{m,(n} a_p a_{q)} \right) \Big[ \alpha^*_{mi} \alpha_{nj} \alpha_{pk} \alpha_{ql} \, \delta_{i, j+k+l} + \beta^*_{mi} \beta_{nj} \beta_{pk} \beta_{ql} \, \delta_{i,j+k+l}  \\ &\quad- 3 \alpha_{mi}^* \alpha_{nj} \alpha_{pk} \beta_{ql} \left( \delta_{i+l,j+k} + \delta_{i,j} \delta_{k,l} + \delta_{i,k} \delta_{j,l} \right) + 3 \beta^*_{mi} \alpha_{nj} \alpha_{pk} \beta_{ql} \, \delta_{i+j+k,l} \\ &\quad- 3 \alpha^*_{mi} \alpha_{nj} \beta_{pk} \beta_{ql} \left( \delta_{i+k,j+l} + \delta_{i,j} \delta_{k,l} + \delta_{i,l} \delta_{j,k} \right) - 3 \beta^*_{mi} \alpha_{nj} \beta_{pk} \beta_{ql} \left( \delta_{i+j,k+l} + \delta_{i,k} \delta_{j,l} + \delta_{i,l} \delta_{j,k} \right) \Big] \\
&- 4 a_m a_n a_p a_q \Big[ \alpha_{mi} \alpha_{nj} \alpha_{pk} \beta_{ql} \, \delta_{i+j+k,l} - 3 \alpha_{mi} \alpha_{nj} \beta_{pk} \beta_{ql} \left( \delta_{i+j,k+l} + \delta_{i,k} \delta_{j,l} + \delta_{i,l} \delta_{j,k} \right) \\ &\quad+ \alpha_{mi} \beta_{nj} \beta_{pk} \beta_{ql} \, \delta_{i, j+k+l} \Big] + H.c.
\end{align*}
Here, we introduce the short notation for the symmetrized quantities, e.g., $A_{(m,n)} = \frac{1}{2!} \left( A_{mn} + A_{nm} \right)$. We also neglect rapidly oscillating and constant terms that do not contribute to~\eqref{eq:n} or~\eqref{eq:k} in the limit in question. In other words, we average the Hamiltonian, $H_\text{int}(t) \longrightarrow \frac{1}{T} \int_t^{t+T} H_\text{int}(t') dt'$, and neglect the subleading contributions in the limit $T \to +\infty$.

It is worth mentioning that the Kronecker deltas of the form $\delta_{i+j,k+l}$ appear only in a massless theory. In a massive two-dimensional theory, these deltas are multiplied by the time-dependent function, $\frac{\sin[ (\omega_i + \omega_j - \omega_k - \omega_{i+j-k}) T]}{(\omega_i + \omega_j - \omega_k - \omega_{i+j-k}) T}$, where $\omega_k = \sqrt{ ( \pi k / \tilde{\Lambda} )^2 + m^2}$. This function goes to one only if $i=k$, $i=l$, or $m \to 0$; otherwise, it vanishes at large evolution times, $T \gg 1/m^2 \tilde{\Lambda}$. This behavior illustrates the well-known fact that elastic scattering of massive particles in a flat two-dimensional spacetime reduces to a simple momentum exchange.

Finally, substituting the approximate Bogoliubov coefficients~\eqref{eq:B} into this identity and discarding the subleading in $\epsilon$ terms, we straightforwardly obtain Eq.~\eqref{eq:H}.

\section{The Hamiltonian of a Josephson metamaterial}
\label{sec:Josephson}

Following~\cite{Blais} and~\cite{Lahteenmaki}, we straightforwardly obtain the Hamiltonian of a Josephson metamaterial that consists of an array of dc-SQUIDs (Fig~\ref{fig:circuit}):
\begin{figure}[t]
    \center{\includegraphics[scale=0.5]{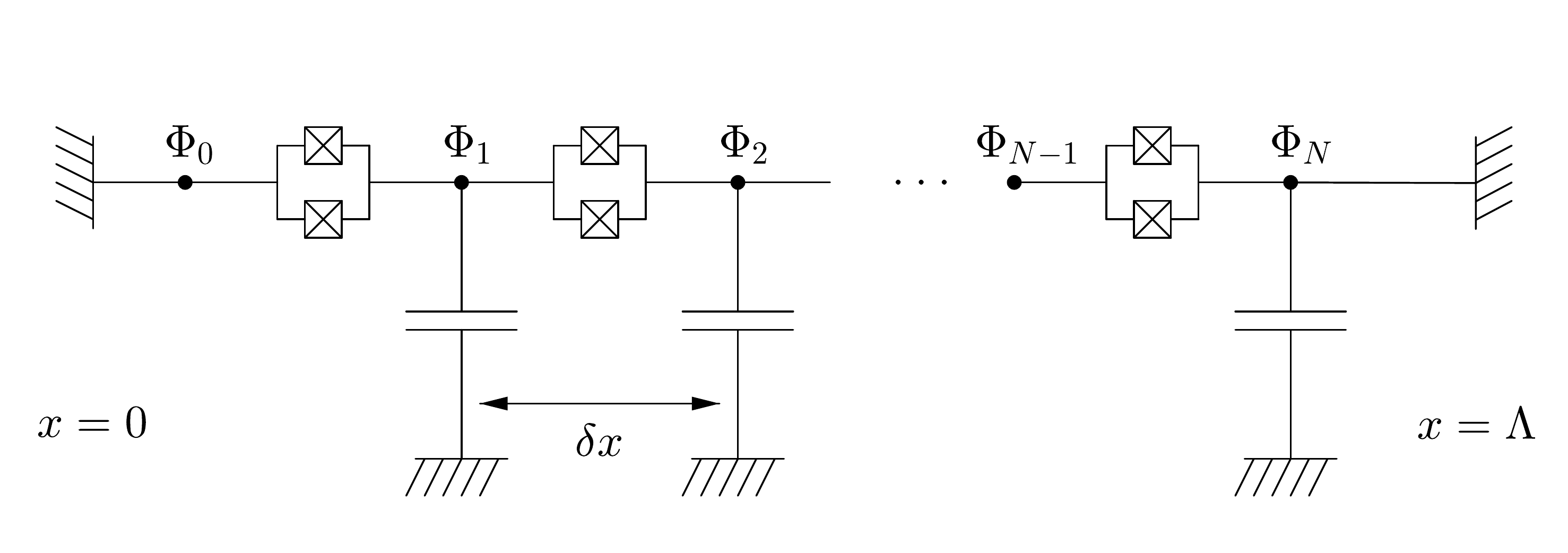}}
    \caption{A schematic representation of the Josephson metamaterial~\cite{Lahteenmaki}, which models the DCE in a cavity with a varying speed of light. The square with a cross represents a Josephson junction.}
    \label{fig:circuit}
\end{figure}
\beq H_{JM} = \sum_{n=1}^N \left\{ \frac{1}{2 C_0} Q_n^2 + E_J(\Phi_\text{ext}) - E_J(\Phi_\text{ext}) \cos \left[ \frac{2 \pi}{\Phi_0} \left( \Phi_n - \Phi_{n-1} \right) \right] \right\}. \eeq
Here, $\Phi_n(t) = \int_{-\infty}^t V_n(t') dt'$ is the flux variable associated with the $n$-th node, $Q_n$ is the charge at that node, $\Phi_0 = \pi \hbar/e$ is the magnetic flux quantum, $C_0$ is the capacitance, and $E_J$ is the Josephson energy of each SQUID. In an external magnetic field, the Josephson energy varies, $E_J(\Phi_\text{ext}) = 2 E_J(0) \cos \left( \frac{\pi \Phi_\text{ext}}{\Phi_0} \right)$ (we assume that both Josephson junctions of the SQUID have approximately the same properties). The total length of the array is $\Lambda$, and the distance between the adjacent nodes is fixed, $\delta x = \Lambda/N$. Note that the array is grounded at both ends, which models the Dirichlet boundary conditions for propagating excitations.

Now let us derive the continuum limit of this Hamiltonian. To that end, we introduce the capacitance and charge per unit length, $c_0 = C_0/\delta x$ and $Q(t,x_n) = Q_n(t)/\delta x$, use the identity $Q_n(t) = C_0 \pd_t \Phi_n(t)$ and expand the cosine to the fourth order:
\beq \label{eq:HJM}
\begin{aligned}
H_{JM} &\approx \sum_{n=1}^N \delta x \left[ \frac{c_0}{2} \left(\pd_t \Phi_n \right)^2 + \frac{1}{2 l_0} \left( \frac{\Phi_n - \Phi_{n-1}}{\delta x} \right)^2 - \frac{1}{24 l_0} \left( \frac{2 \pi \delta x}{\Phi_0} \right)^2 \left( \frac{\Phi_n - \Phi_{n-1}}{\delta x} \right)^4 \right] \\ &\approx \int_0^\Lambda dx \left[ \frac{1}{2} (\pd_t \phi)^2 + \frac{1}{2} v^2 (\pd_x \phi)^2 - \frac{\lambda v^2}{4} (\pd_x \phi)^4 \right].
\end{aligned} \eeq
For shortness, we also introduce the inductance per unit length, $l_0(t,x)= \frac{1}{\delta x E_J\left[\Phi_\text{ext}(t,x)\right]} \left( \frac{\Phi_0}{2 \pi} \right)^2$, the effective speed of light, $v(t,x) = 1/\sqrt{c_0 l_0(t,x)}$, and the effective coupling constant $\lambda = \frac{1}{6 c_0} \left( \frac{2 \pi \delta x}{\Phi_0} \right)^2$. In the last identity of~\eqref{eq:HJM}, we also redefine the flux variable, $\phi(t,x) = \sqrt{c_0} \Phi(t,x)$. Note that in practice, variations of the external magnetic field are usually spatially uniform, $\Phi_\text{ext}(t,x) = \Phi_\text{ext}(t)$. However, we consider arbitrary functions $\Phi_\text{ext}(t,x)$ for generality.

Now it is easy to see that in the naive limit $\lambda \to 0$, which is justified at small evolution times, $t \ll \Lambda^3/\hbar \lambda$, Josephson metamaterial effectively reproduces the standard model of the scalar DCE, Eq.~\eqref{eq:DCE}. Otherwise, Hamiltonian~\eqref{eq:HJM} implies the equations of motion~\eqref{eq:DCE-squid}.

\end{document}